\documentclass[twoside,reqno]{HERON}
\usepackage{epsfig,cite,colordvi}
\usepackage{graphicx}
\usepackage{url}
\usepackage{amssymb,amsmath,amscd,epsf}
\usepackage{times}
\usepackage{makeidx}
\def\n{\noindent}
\def\[{[\![}
\def\]{]\!]}

\def\l{\ldots}

\def\o{\omega}

\def\P{{\bf P}}
\def\R{{\bf R}}

\def\hP{{\hat P}}
\def\hR{{\hat R}}
\def\hp{{\hat p}}
\def\hq{{\hat q}}

\def\hbP{{\hat {\bf P}}}
\def\hbR{{\hat {\bf R}}}

\def\T{\Theta}
\def\t{\theta}

\def\d{\delta}
\def\i{{\rm i}}
\def\ra{\rangle}

\begin{document}
\title{DO NEW QUANTUM STATISTICS EXIST IN NATURE?}

\runningheads{DO NEW QUANTUM STATISTICS EXIST IN NATURE?}
{T.D. Palev}..

\begin{start}

\author{T.D. Palev}{1}

\index{Palev T.D.}

\address{Institute for Nuclear Research and Nuclear Energy, 1784 Sofia,
Bulgaria}{1}


\textbf{Abstract}. We recall the definitions  
of a Wigner quantum
system (WQS)  and of  $ A-$, 
(resp $B-$, $C-$ and $D-$)
 quantum  (super)statistics.
We  outline shortly the relation of these 
new statistics to the  classes $A$, 
(resp $B$, $C$ and $D$) of Lie (super)algebras. 
We describe in some more
details some of the properties 
of  $A-$oscillator and $A-$superoscillator. Both of them fall 
into the category of finite quantum systems 
and  both of them have quite unusual 
properties. For example let $\hR$ and
$\hP$ be the position and the momentum
operators of an 3D $A-$ superoscillator 
along the $x$  (or $y$ or $z$) axes. Then 
contrary to the canonical uncertainty relations
the product  of the standard deviations
of $\hR$ and $\hP$  of the oscillating particle reads:
$$
\Delta \hR .\Delta \hP \leq  {\hbar/2}.  
$$ i.e., the inequality above $\le$  is
 just the opposite to the one that appears in the
 canonical quantum mechanics.


\end{start}

\section{Introduction. Wigner Quantum Systems}\label{sec1}



In \cite{Thesis}  we have shown that the canonical  quantum 
statistics, i.e., the Bose and the Fermi statistics,
can be considerably generalized  if one abandons the 
requirement that the commutators or the 
anticommutators between the fields in quantum field
 theory (QFT) or the commutators
between the position and the momentum operators
 in quantum mechanics (QM) to be $c-$numbers.

The purpose of the present paper  is to outline shortly
where the idea (we refer to it as to
{\it  the main idea})
for  a possible 
generalization of quantum statistics 
came from,  and to list as an illustration some 
of the  unusual predictions of the new statistics,
 mainly of $A-$(super)statistics. Throughout we 
skip the proofs of most
of the propositions,  which  will considerably 
simplify the exposition.


Chronologically Wigner was the first, who, back in 1950, made a
decisive step towards generalization of quantum statistics in
 quantum mechanics \cite{Wigner}. In
order to indicate where this possibility came from, we first recall
the axioms of quantum mechanics as given by Dirac \cite{Dirac}

\bigskip\n
A1. To every state of the system there corresponds a normed 
to 1 wave function           $\Psi$.

\n
A2. To every physical observable $L$ there corresponds 
a selfadjoint operator $\hat L$.

\n
A3. The observable $L$ can take only those values, which 
are eigenvalues of
 $\hat L$.
                        
\n
A4. The expectation value $ L_\Psi $ of $L$ in the state $\Psi$ is
 $\langle\Psi|\hat L|\Psi \rangle$.

\n
A5. The Heisenberg  equations in the Heisenberg picture  hold:
$$
{\dot{\hat p}}_k =
-{\i\over {\hbar}}[{\hat p_k, H]},~~~ {\dot{\hat q}}_k =
-{\i\over {\hbar}}[{\hat q_k, H]}.                                          
     \eqno(1)
$$
A6.  The canonical commutation relations  (CCR's) hold:
$$
[{\hat q}_j ,{\hat p}_k]={\rm i}\hbar\d_{jk},~
~[\hq_j,\hq_k]=[\hp_j,\hp_k]=0.    \eqno(2)
$$
A key outcome from the above axioms and especially 
from A5 and A6 are
the following consequences:

\n
C1.  The (operator) equations of motion
 (the Hamiltonian equations) hold too,
$$
{\dot{\hat p}}_k = 
 -{{\partial H}\over{\partial {\hat q}_k}},
~~~
 {\dot{\hat q}}_k = 
 {{\partial H}\over{\partial {\hat p}_k}}.                                                              \eqno(3)
$$
C2. From the CCR A6 and the equations of motion C1 one derives
the Heisenberg equations C2.

At this place let us ask a question: can some of the
 axioms A1-A6 be removed or modified or replaced or
 be weakened somehow?

Clearly no one from the first four axioms can be removed 
without
changing the very essence of QM. For the same reason 
one cannot
touch A5, since in the Schroedinger representation this 
axiom leads to the Schroedinger equation! It is also clear
that the equations of motion (C1)  have to hold in
any case,
since they are responsible for the correct classical limit. 

 What is left? 
It remains to analyze axiom A6. The first impression 
is  that the axiom A6 has to remain too, since 
then also the  Hamiltonian  equations  C1 hold. The more
precise statement is however that the  CCR's are sufficient
in order C1 to be fulfilled. Are the CCR's also necessary? This
as a question asked by Wigner \cite{Wigner}. And the
 answer was "NO"! The CCR's
are not necessary in order  the equation of motion 
C1 to hold!
Wigner proved this \cite{Wigner}
on an example of one-dimensional
harmonic oscillator. Unknown were the position and 
the momentum operators. Parallel to the canonical 
 solution  Wigner found
 infinitely many other solutions, which also satisfied
 A5 and C1, but not A6.  
      
Having observed this Wigner remarked that from a 
physical point of view
the Heisenberg equation and the Hamiltonian equations 
have a more direct physical
significance than the CCR's.  Therefore it is justified  
to postulate from the very beginning that the equations
 C1 hold instead 
of the CCRs. Based on all this we introduce

\n {\bf Definition 1.} {\it A quantum system subject 
 to  axioms A1-A5 and   C1 is said to be a Wigner Quantum 
System  (WQS). A WQS is noncanonical if it does not 
satisfy A6, i.e. the canonical commutation relations.}
 
This definition was introduced for the first time 1982 in
 \cite{Sashka}.

The statistics of canonical quantum mechanics is  
"hidden" in axiom A6, since  the related
creation and the annihilation operators (CAOs)
$$
b_i^+ = {1\over{\sqrt 2}}(q_k - i p_k),~~{\rm and}~~
b_i^- = {1\over{\sqrt 2}}(q_k + i p_k)              \eqno(4)  
$$
obey the Bose commutation relations:
$$
[b_i^- , b_j^+]=\d_{ij} ,~~[b_i^- , b_j^-]
=[b_i^+ , b_j^+]=0.  \eqno(5) 
$$

Since the left hand sites of  A5 and C1 coincide, their 
right hand sites have to
coincide too. Therefore the definition of a WQS is 
selfconsistent if 
 {\it the  main quantization condition} holds, namely:
$$
{\i\over {\hbar}}   [{\hat p_k, H]}
= {{\partial H}\over{\partial {\hat q}_k}}              \eqno(6a) 
$$
$$
-{\i\over {\hbar}}[{\hat q_k, H]} =
 {{\partial H}\over{\partial {\hat p}_k}}     \eqno(6 b)  
$$
These two equations actually correspond to the quantization in QM.
Note that they depend on the Hamiltonian. In fact we 
know that these equations have a solution for  Hamiltonians,
corresponding to harmonic oscillators. What are
 the admissible Hamiltonians 
in the general case is an
open question. This is hardly a surprise. In quantum
 field theory the harmonic oscillator 
 interactions correspond to free fields interactions and so far one knows how to quantize rigorously
 only  free fields.

\section{B-statistics  }\label{sec2}

The next big step towards generalization of quantum statistics
is due
to Green \cite{Green }, 
who in 1953  discovered the
paraFermi (pF)  and the paraBose (pB) 
statistics as new possible statistics in the free quantum field theory (QFT). 
In this more general setting the Fermi anticommutation relations
were replaced by  the more general double commutation relations for
paraFermi CAO's, namely:
$$
[[f_i^+, f_j^-],f_k^+]=2\d_{jk}f_i^+,                                            
$$ 
$$
[[f_i^+, f_j^-],f_k^-]= - 2\d_{ik}f_j^-,   \eqno(7)                                 
$$ 
$$ 
[[f_i^+, f_j^+],f_k^+]= [[f_i^-, f_j^-],f_k^-]= 0.                   
$$ 
Similarly the Bose  commutation relations were  replaced by similar relations, but part of the
commutation relations have to be replaced by anticommutation relations:
$$
[\{b_i^+, b_j^-\},b_k^+]=2\d_{jk}b_i^+,                                             
$$ 
$$
[\{b_i^+, b_j^-\},b_k^-]= - 2\d_{ik}b_j^-,      \eqno(8)                               
$$ 
$$ 
[\{b_i^+, b_j^+\},b_k^+]= [\{b_i^-, b_j^-\},b_k^-]= 0.                    
$$

From the above triple relations  there  comes for the first time an indication about  possible  connection between  quantum statistics and Lie (super)algebras \cite{KR}.

Indeed, it is almost evident  from (7)
that the linear envelope 
$$
lin.env.\{f_i^\xi,  [f_j^\eta, f_k^\epsilon] ~  i,j,k=1,....,n, \xi, \eta, \epsilon =+, - \}   \eqno(9)    
$$ 
of all CAOs and their commutators is a Lie algebra(LA). It takes some time to prove that this LA is the   algebra of the orthogonal group  $so(2n+1)$ denoted  also as $B_n$, if in (7) $i,j,k=1,2,\ldots,n$. Thus, the parafermi quantization corresponds to a case when any $n$ pairs of position and the momentum operators are elements from the Lie algebra $B_n$ and they generate $B_n$ . For this reason  it is natural to call the paraFemi statistics also $B-$statistics.
 The circumstance that the paraFermi operators have to satisfy
the triple relations (7)  means actually that these operators determine  a
 representation of $so(2n+1)$.

{\bf Definition 2.} {\it Let $W$ be a WQS. If its position and momentum operators (PMO's) generate  (a representation of)  a Lie algebra $G$, then the corresponding quantium system is said to be a $G-$ quantum system, the related quantization  - a $G$-quantization and the related statistics a $G-$statistics.}

In the very definition of paraFermi statistics it is required that the representations are of 
Fock type \cite{Green}. This is achieved from the requirement that the state space contains a
vacuum vector, so that 
$$
f_i^-f_j^+ |0\ra = \d_{ij}p |0\ra,                    \eqno(10)  
$$ 
where $p$ is a positive integer. The number $p$ is said to be  
{\it an order of statistics} \cite{Greenberg}.  It labels the different representations. The representations corresponding to different $p$
 are inequivalent. The pF statistics with an order of statistics $p=1$ 
 is the  Fermi statistics.

As mentioned above the order of statistics is a positive integer. The representations with noninteger order of statistics do also
exist and in certain cases could be of interest too \cite{P2}

The transformations of the Fock space under the action of the paraFermi CAO's is  not simple and is in fact very difficult task. For more details we refer to\cite{Ohnuki}. The main difficulty stems from the
observations that the paraFermi creation operators do not commute with each other. In view
of this the very definition of the Fock space  and its interpretation is nontrivial \cite{Green},\cite{Ohnuki}.

 Looking into some more mathematical books on Lie algebras, we learn that the algebra $B_n$
 belongs to the class  $B$ of  the only  four infinite classes of simple LA
$A$, $B$,  $C$ and  $D$. 

Based on   the above observations we  draw the following

\n {\bf Conclusion 1.} {\it The paraFermi quantization  corresponds 
 to quantization with
position and momentum operators
 (or to the related creation and annihilation operators)
which generate a Lie algebra from the class $B$.}

For this reason we give 

\n {\bf Definition 3.} {\it We call the paraFermi statistics also a $B-$statistics and the related
quantization a $B-$quantization }.

The above conclusion rises an immediate 

\n {\bf Question 1.} {\it  Do there exist quantum statistics with position and momentum operators 
 (or creation and annihilation operators) ,which generate Lie algebras from the classes $A$,
$C$ or $D$.}

A positive answer to the above question  was given in\cite{Thesis}. 

\n {\bf Definition 4.}  {\it The statistics, which
 position and momentum operators 
 (or creation and annihilation operators)  generate Lie algebras from the classes 
 $A$, $C$ or $D$ are called $A-$,
$C-$ or $D-$statistics, respectively.}

Let us now turn and analyze shortly the algebraic structure of
paraBose (pB) operators. The defining relations for pF and pB operators (7) and (8)
look pretty similar apart from the circumstance that part of the commutators in (7) are replaced by anticommutators in (8). 
This difference however turns to be  essential. It is clear  what we are searching for.
We  have  to find all operators (8), i.e., we have to determine
all representations of the relations (8) and then select those of them, which obey the restrictions of QM.  How to determine 
the representations?  The idea is to reduce the
unknown problem to a more or less known one, similar as we did for 
paraFermi operators. To this end we  first introduce appropriate 
terminology and notation.

To begin with we define two subspaces, called an even one  $ B_0 $
and   an odd  one $ B_1$, namely
$$
B_0 = lin.env.\big\{ \{b_i^\pm,b_j^\pm \} | ~i,j=1,2,...,n, \big\}
$$
$$ 
B_1 = lin.env.
\big\{ b_k^\pm | k=1,2,...,n. \big\}
 \eqno(11)
$$

Let  ${\bf B}_n$ be their direct sum (in the sense of linear spaces)
$$
  {\bf B}_n=B_0 \oplus B_1. \eqno(12)
$$
The  elements from $ B_0$ (resp  $B_1$)
are  also said to be   {\it even (resp odd) elements}.
Define a supercommutator   $\[x,y\] $ on ${\bf B_n}$, setting
$$
\[x,y\] = \{x,y \},~~{\rm if}~~x,y~~{\rm are~both~odd }, \eqno(13)
$$
$$
\[x,y\]= [x,y], ~~{\rm if}~~x~{ \rm or}~y~{\rm or~both}~
{\rm are~even}~.   \eqno(14) 
$$
and extend the above relations to every two elements by linearity.
Then by definition ${\bf B}_n$
is a Lie superalgebra (LS) and the problem to determine all 
 paraBose operators  reduces to the task to determine the representations of the 
LS ${\bf B}_n$.

Here we had luck! Without
 going into  the   details, because we do not need them, we mention that ${\bf B}_n$ belongs to the class $\bf B$ of basic Lie superalgebras \cite{Kac} and therefore
the paraBose statistics can be called ${\bf B}-superstatistics$. There are
four classes of 
basic Lie superalgebras in the classification of Kac \cite{Kac}, namely the classes  ${\bf A}$, ${\bf B}$, ${\bf C}$ and ${\bf D}$. 
Hence one can ask whether it is possible to associate
creation and annihilation operators with each class of basic  LS's
and talk about 
${\bf A-}$, ${\bf B-}$, ${\bf C-}$ and ${\bf D-statistics}$,  respectively

Finally, 
each Lie algebra is a Lie superalgebra with only even elements.  
Moreover it turns out that each class of simple Lie algebras is a subset of
the corresponding class of basic Lie superalgebras, namely
$$
 A\subset {\bf A}, ~ B\subset {\bf B},~  C\subset {\bf C},~  D\subset {\bf D}.   \eqno(15)
$$
Therefore,
it is possible to unify the Lie algebra case with Lie superalgebra one,
 setting:

\n {\bf Question 2.} {\it  Do there exist quantum statistics with position and momentum operators 
 (or creation and annihilation operators) ,which generate 
Lie (super)algebras from the classes $\bf A$,  $\bf B$,  $\bf C$, or
$\bf D$ ?}

The answer to the above question is positive, but not complete. So far 
we have studied in more details the statistics, corresponding to the class $A$ Lie algebras and Lie superalgebras. Let us say some more 
words  about $\bf A$-(super)statistics.

\section{A-statistics  }\label{sec2}

\bigskip\n 
The CAO's of $A$-statistics satisfy  the  relations:
$$
 [[a_i^+,a_j^-],a_k^+]=\delta_{jk}a_i^+ +\delta_{ij}a_k^+, 
$$
$$
[[a_i^+,a_j^-],a_k^-]=-\delta_{ik}a_j^- -\delta_{ij}a_k^-, 
\eqno(16)
$$
$$
[a_i^+,a_j^+] = [a_i^-,a_j^-] =0.      
$$

Here we
review shortly some of the properties of these CAO's and of
their
Fock representations.

The first remarkable property of $A$-statistics is that the creation operators commute with each other. This simplifies greatly all computations of $A$ statistics and in particular the construction
and the interpretation of the Foch space.

As indicated,  the
 name $A$-statistics comes to remind that the operators
$a_1^\pm,...,a_n^\pm$ (and more generally any $n$ different 
pairs of $A$-CAO's) 
generate a Lie algebra $A_n \equiv sl(n+1)$ from the class
${\cal A}$.
In particular a set of $sl(n+1)$ generators, which constitute a 
linear basis  in the underlying
linear space, can be taken to be ($i\ne j=1,\l,n$):
$$
e_{i0}=a_i^+, ~e_{0i}=a_i^-, ~ e_{ii}-e_{00}=[a_i^+,a_i^-],~
e_{ij}=[a_i^+,a_j^-]. \eqno(17)
$$

Above $\{e_{ab}|a,b=0,1,\ldots,n\}$ are 
generators of $gl(n+1)$:
$$
[e_{ab},e_{cd}]=\delta_{cb}e_{ad}-\delta_{ad}e_{cb}.\eqno(18)
$$

By definition  each Fock space\cite{P3}   is an irreducible
$sl(n+1)-$module, which 
satisfies certain natural for physics
requirements, namely the metric  in each Fock space
$W$ is positive definite, 
$W$ contains an
unique vector $|0\ra$, called a vacuum, so that $a_i^-|0\ra=0$
and the
Hermitian conjugate to $a_k^-$ is $a_k^+$ (for any integer $k$), i.e., 
$
(a_k^+)^\dagger=a_k^-.
$
It is proved \cite{P3} that  the irreducible inequivalent
Fock spaces   $W_p$
are labeled by all positive integers $p=1,2,...$, the order of statistics. 
Each  Fock space $W_p$ is a finite-dimensional 
irreducible state space. All vectors
$$
|p;l_1,..,l_n)=\sqrt{(p-\sum_{j=1}^n l_j )!\over p!l_1!l_2!\l
l_n!}
(a_1^+)^{l_1}\l(a_n^+)^{l_n} |0\ra, \eqno(19)  
$$
subject to the condition $ l_1+\l+l_n\le p$
constitute an orthonormal basis in $W_p$. The transformations of
the basis (19)
under the action of the CAOs read:
$$
a_i^+|p;l_1,.,,l_i,..,l_n)=
  \sqrt{(l_i+1)(p-\sum_{j=1}^n l_j )}~
|p;.,l_i +1,..,l_n),    \eqno(20a)                   
$$
$$
 a_i^-|p;l_1,..,l_i,..,l_n)=
  \sqrt{l_i(p-\sum_{j=1}^n l_j +1  )}~
|p;l_1,.,l_i-1,.,l_n).        \eqno(20b)              
$$

In a consistent with (18) way we extend $W_p$ to an irreducible
$gl(n+1)$-module, setting for the central element
$$
e_{00}+e_{11}+... +e_{nn}=p.    \eqno(21)  
$$
 Then     

$$
 e_{00} |p;l_1,l_2,...,l_n)=(p-\sum_{i=1}^nl_i)|p;l_1,l_2,...,l_n),
\eqno(22)
$$
$$
e_{ii}|p;l_1,l_2,...,l_n)=l_i|p;l_1,l_2,...,l_n), ~~ i=1,..,n. \eqno(23)
$$
  
The operators $e_{00}, e_{11},...,e_{nn}$ commute. From (23) we
conclude that
$e_{ii}$ can be interpreted as a number operator for the
particles on the
orbital $i$.
Then $|p;l_1,\l l_{i-1},l_i,l_{i+1},\l,l_n\ra$ is a state
containing $l_1$
particles on orbital $1$, $l_2$ particles on orbital $2$, and so
on,..,$l_n$
particles on orbital $n$. 
Note that for a given $p$,  
$$
(a_1^+)^{l_1}\l(a_n^+)^{l_n}|0\ra =0,~~{\it if}~~
l_1+l_2+...+l_n> p.           \eqno(24)  
$$
One immediate  conclusion from (24) is evident:

\n {\bf Corrolary A.} { \it For any $p$ the state
space $W_p$ is a finite-dimensional linear space.}

\bigskip\n
The second conclusion from (24)  is
 actually the

\bigskip
\n{{\bf Pauli principle for $A-$statistics}: {\it If the order of
statistics is $p$,
then each basis state $|p;l_1,l_2,...,l_n)$ from $W_p$
corresponds to
$l_1+l_2+...+l_n \le p$ particles. There are no states with more
than $p$
particles in $W_p$.}

 This issue holds certainly only in
a particle interpretation of the picture.

\smallskip
 Another relevant property of
$A-$statistics is that  in the limit $p \rightarrow \infty $  the operators
$$
b(p)_k^\pm = {a(p)_k^\pm /{\sqrt p}},    \eqno(25)
$$
are becoming ordinary bosons.

We will not try to interpret a system of $A$ particles as a WQS. From what we have stated above  it is clear that the particles of $A-$statistics can be interpreted perfectly well as quasiparticles. In the next example however we will show how a real 3D harmonic  oscillator  can be quantized with $A-$superstatistics into a WQS with 
very unusual for QM properties.
 
\bigskip
              
\section{A-superstatistics  }\label{sec2}

\bigskip\n 
Here we shall  indicate how one can quantize a $3D$
harmonic oscillator based on 
$A-$superstatistics and more precisely with the  CAO's of the
LS $sl(1|3)$.

 The creation and the annihilation operators of $A-$superstatistics, and more precisely
of $sl(1|n)$  (with $i,j,k=1,2,...,n$ bellow) read:
$$
 [\{a_i^+,a_j^-\},a_k^+]=\delta_{jk}a_i^+ - \delta_{ij}a_k^+, 
\eqno(26a)
$$
$$
 [\{a_i^+,a_j^-\},a_k^-]=-\delta_{ik}a_j^- +\delta_{ij}a_k^-, 
\eqno(26b),
$$
$$
  \{a_i^+,a_j^+\} = \{a_i^-,a_j^-\} =0,    \eqno(26c)  
$$
The  linear span of all CAO's yields the odd subspace and the linear combinations
of all of their anticommutators is the even subspace.
In the   case $i,j,k=1,2,3$ the CAO's generate the LS $sl(1|3)$.

The Fock representations, which we consider, are determined from the requirements
that there exists a vacuum vector  $|0\ra$, so that
$$
a_i^-|0\ra =0,~~ a_i^-  a_j^+|0\ra=\d_{ij}p|0\ra, ~~ i,j=1,2,3. 
\eqno(27)  
$$
Moreover the hermiticity condition  $(a_i^+)^* = a_i^-$   with ($*$)  being hermitian conjugation holds.

To each  positive  integer  $p$, called {\it an order of statistics,}  there corresponds
an irreducible representation space $W(p)$. The representations corresponding to different $p$
are inequivalent.

Let $\T\equiv \t_1, \t_2, \t_3$ with $\t_i\in (0,1)$ for any $i=1,2,3.$              
Define an orthonormed   basis in $W(p)$ :
$$
|p,\T\ra \equiv |p;\t_1,\t_2,\t_3\ra =
 {\sqrt{(p-q)!\over p!}} (a_1^+)^{\t_1} (a_2^+)^{\t_2}   (a_3^+)^{\t_3}|0\ra ,              \eqno(28)
$$
where  
$$
0 \le q \equiv \t_1 +\t_2 +\t_3 \le min(p,3),                                                                       ~~ \t_1, \t_2, \t_3 = 0,1.
$$
The transformations of the basis under the action of the CAO's reads:
$$
a_i^- |p;...,\t_i...\ra =\t_i (-1)^{\t_1 +...+\t_{i}} \sqrt {p-q+1}|p;...,\t_i-1,...\ra,          \eqno(29a)
$$
$$
a_i^+ |p;...,\t_i...\ra =(1-\t_i) (-1)^{\t_1 +...+\t_{i}} \sqrt {p-q}|p;...,\t_i+1,...\ra.      \eqno(29b)    
$$

We proceed to show that any  (3D) harmonic oscillator can be quantized
with the CAO's of $A-$ superstatistics
defined above. This means that  the physical observables related to the 
oscillator can be expressed via 
the CAO's of $A-$superstatistics, so that the oscillator will be a Wigner quantum system.

The Hamiltonian 
$$
H={\hbP^2\over 2m} + {m\o^2\over 2}\hbR^2 ,  \eqno(30)
$$
the equations of motion
$$
{\dot{\hbP}}=-m \o^2 {\hbR}, ~~{\dot{\hbR}}={1\over m} {\hbP}
 \eqno(31)
$$
and the Heisenberg equations
$$
{\dot{\hbP}}={i\over \hbar}[H, {\hbP}],~~ 
{\dot{\hbR}}={i\over \hbar}[H, {\hbR}]   \eqno(32)
$$
are well known. For the compatibility equations we have
$$
[H, {\hbP}]=i\hbar  m \o^2 {\hbR},~~[H, 
{\hbR}]=-{i\hbar\over m}{\hbP}.   \eqno(33)
$$

In order to express the Hamiltonian, the position and the momentum operators via CAO'
we set:
$$
a_k^\pm = \sqrt{m \o \over  2   \hbar} R_k \pm i \sqrt{1\over 2m \o \hbar}P_k, ~~k=1,2,3.   \eqno(34)
$$

Then the Hamiltonian and the compatibility relations read:
$$
H={\ \o \hbar\over 2} \sum_{i=1}^3 \{  a_i^+,a_i^- \}  \eqno(35)
$$
$$
\sum_{i=1}^3 [\{  a_i^+,a_i^- \}, a_k^\pm]=\mp 2 a_k^+ ,~~i,k=1,2,3.       \eqno(36)
$$
Observe that in the derivation of the above relations we have not used
the triple relations (26). Taking now into account the triple relations we immediately
see that  the compatibility equations are satisfied. The solution of the  Heisenberg and the Hamiltonian 
equations are:

$$
 \hR(t)_k ={ \sqrt{\hbar \over{2m\o}}} (a_k^+ e ^{-i\o t} + a_k^-{e ^{i\o t}} ),~     \eqno(37)
$$
$$
 \hP(t)_k ={ - i \sqrt{\hbar m \o \over{2}}} (a_k^+ e ^{-i\o t} + a_k^-{e ^{i\o t}} ),~    \eqno(38)
$$
From all stated above we conclude that the oscillator has solutions as a 
Wigner quantum system. Let us  mention some  of its properties.

 On the first place,
contrary to the canonical oscillator,  each irreducible  state space $W(p)$ 
has no more than four equally spaced energy levels with spacing $\o \hbar $,
more precisely,
$$
H|p; \t_1,\t_2,\t_3\ra={\o \hbar\over 2} (3p-2q) |p; \t_1,\t_2,\t_3\ra,~~
 q=\t_1+\t_2+\t_3=0,1,...,min(p, 3).   \eqno(39)
$$
The next relation gives more information:
$$
{2\over {\o \hbar}}H = {2m\o\over \hbar}\hbR^2=
{2\over{m\o\hbar}}\hbP^2=
\sum_{i=1}^3\{a_i^+,a_i^- \} \eqno(40)
$$
 Therefore $H,\R^2$ and $\P^2$            commute with each other. We say that the geometry is square commutative, but not commutative.         
In particular 
$$
\hbR^2|p;\T\ra={\hbar\over{2m\o}}(3p-2q)|p;\T\ra.     \eqno(41)
$$
$$
\hbP^2|p;\T)={m\o \hbar\over 2}(3p-2q)|p;\T)             \eqno(42)
$$
Therefore if $p>2$, then $\R^2$ for instance can take no more then four different values. This means that the distance of the particle with respect to the center of the coordinate system is fixed and has no more than four values. One can be even more precise taking into account that all operators 
$$
H,  \hbR^2, \hbP^2, \hR_1^2, \hR_2^2,  \hR_3^2, \hP_1^2, \hP_2^2, \hP_3^2,        \eqno(43)
$$
commute, which is not the case in canonical quantum mechanics. If
 the oscillator is in a state $|p;\Theta\ra$ with $p>2$, then
$$
\hR_i^2|p;\theta_1,\theta_2,\theta_3\ra=\frac{\hbar}{2m\o}(p-q+\theta_i) |p;\theta_1,\theta_2,\theta_3\ra, \;\; i=1,2,3.    \eqno(44)
$$
$$
\hP_i^2|p;\theta_1,\theta_2,\theta_3\ra=\frac{m\o\hbar}{2}
(p-q+\theta_i) |p;\theta_1,\theta_2,\theta_3\ra, \;\; i=1,2,3.    \eqno(45)
$$

The conclusion is that if the particle is in a state $|p;\Theta>$, with $p>2$, then at each measurement it can be spotted in no more than 8 different points of the (3D) space with coordinates along $x,y, z$ axis in units $\sqrt{\frac{\hbar}{2m\o}}$ as follows:
$$
x=\pm\sqrt{p-q+\theta_1}, \;\; y=\pm\sqrt{p-q+\theta_2}, \;\; z=\pm\sqrt{p-q+\theta_3}.\eqno(46)
$$ 

We resume the present exposition by 
pointing out  that for the superoscillator
 under  consideration
the uncertainty relations are not fulfilled.  

We begin the proof with the observation 
that if the oscillator is in any basis
state $|p;\T\ra$, then the average value of its coordinates and 
momenta  vanish. i.e.
$$
\bar{R}_i=0,  \bar{P_i}=0, i=1,2,3. \eqno(47)
$$

Let the oscillator be in a  basis state $|p;\T\ra$.
In view of (47) the expression for the standard deviation 
becomes:
$$
\sqrt { (|p;\t_1, \t_2, \t_3\ra, R_i^2|p;\t_1, \t_2, \t_3}\ra),
$$
where $(x,y)$ denotes  the scalar product between the vectors $x,y$
from the state space of the oscillator. Applying the result from  (44),
we obtain.
$$
\Delta R_i = \sqrt {\frac{\hbar}{2m\o}(p-q+\theta_i) .} 
   \eqno(48) 
$$

Similarly the expression for the standart deviation of the momentum
operators read:
$$
\Delta P_i = \sqrt {\frac{m\o\hbar}{2}(p-q+\theta_i) .}    \eqno(49)
$$

The Eqs. (48) and (49) imply that
$$
 \sqrt {\frac{\hbar}{2m\o}p}\geq
\Delta R_i \geq \sqrt {\frac{\hbar}{2m\o}(p-2)},   
   \eqno(50) 
$$

$$
\sqrt {\frac{m\o\hbar}{2}p}\geq
\Delta P_i  \geq  \sqrt {\frac{m\o\hbar}{2}(p-2)}. \eqno(51)
$$
Hence,
$$
  {p\hbar/2}\geq \Delta R_i .\Delta P_i \geq  {(p-2)\hbar/2}.     \eqno(52)
$$
The above inequality incorporates the canonical kind
uncertainty relation, since it put a restriction from below
on $\Delta R_i .\Delta P_i$. But it restricts also 
$\Delta R_i .\Delta P_i $ from above, which is a new feature
typical for the finite quantum systems.

\section{Conclusions}

We have indicated that from purely theoretical point of view there might exit new quantum statistics. The new statistics are selfconsistent, they lead to interesting predictions still to be investigated.
 The question however, formulated in the title, remains open.

\end{document}